\documentstyle[12pt,epsfig]{article}
\newcommand{\institute}[1]{\parbox{16cm}{%
\centering\normalsize \sl #1}}
\newcommand{\slog}{s_{\rm log}}
\newcommand{\smin}{s_{\rm log,-}}
\newcommand{\smax}{s_{\rm log,+}}
\newcommand{\Slog}{S_{\rm log}}
\newcommand{\half}{{\textstyle{\frac{1}{2}}}}
\newcommand{\ts}{$\times$}

\title{{\vspace{-2cm} \normalsize
\hfill \parbox[t]{4cm}{CERN-TH/95-170\\
                    DESY 95-123\\
                    ITP-Budapest-511\\
                    MS-TPI-95-2\\
                    hep-lat/9506029}\\[7em]}
\bf Interface tension of the electroweak\\
phase transition}
\author{%
F.~Csikor\\[0.5ex]
\institute{Institute for Theoretical Physics, E\"otv\"os University,\\
H-1088 Budapest, Hungary}\\[2ex]
Z.~Fodor%
\thanks{On leave from Institute for Theoretical Physics,
E\"otv\"os University, H-1088 Budapest, Hungary}\\[0.5ex]
\institute{Theory Division, CERN,\\CH-1211 Geneva 23, Switzerland}\\[2ex]
J.~Hein\\[0.5ex]
\institute{Deutsches Elektronen Synchrotron DESY,\\
D-22603 Hamburg, Germany}\\[2ex]
J.~Heitger\\[0.5ex]
\institute{Institut f\"ur Theoretische Physik I, Universit\"at M\"unster,\\
D-48149 M\"unster, Germany}}
\date{}
\begin{document}
\maketitle
\begin{abstract} \noindent
In our nonperturbative lattice investigation we study
the interface tension of the finite-temperature electroweak phase transition.
In this analysis the Higgs mass
has been chosen to be about $35$ GeV.
At the transition point of a finite volume system,
tunnelling between the symmetric and
the Higgs phase takes place.
This phenomenon leads to a splitting of the ground state, which
can be used to determine the interface tension.
The result obtained this way agrees with the result of the two-coupling
method and with the prediction of the perturbative approach.
\end{abstract}

\vfill

CERN-TH/95-170

June 1995

\newpage
\section{Introduction}

At high temperatures the electroweak symmetry is restored. In the early
universe a cosmological phase transition took place
between the high-temperature
--``symmetric''-- and low-temperature --``Higgs''-- phases.
Since the baryon violating processes are unsuppressed at high temperatures,
the observed baryon asymmetry of the universe has finally been
determined at the electroweak phase transition \cite{kuzmin}.

In recent years quantitative studies of the electroweak phase transition have
been carried out by means of resummed perturbation theory
\cite{arnold}--\cite{buchmuller} and lattice Monte Carlo simulations
\cite{bunk}--\cite{karsch}. It has been clarified that
in the SU(2)-Higgs model for Higgs masses ($m_H$) below 50 GeV, the
phase transition is of first order. However, its
strength rapidly decreases for increasing Higgs boson masses.

One of the most important quantities that characterize the strength of the
phase transition is $\sigma$, the interface tension between the two phases.
The course of the cosmological electroweak phase transition, thus the
nucleation rate and temperature, is substantially influenced
by $\sigma$.

The determination of $\sigma$ for $m_H \approx 35$ GeV
in the SU(2)-Higgs model is the main goal of the present letter.
We propose and apply some new methods different from those of
\cite{su2higg_npb}.
Comparing the results we strengthen the confidence in the correctness of
the determination of $\sigma$.

The plan of the paper is as follows.
In Section 2 we give some details of the simulations and suggest a
new method to determine the transition point. Section 3 contains
the determination of $\sigma$ based on the finite volume
tunnelling phenomena between the two phases.
In Section 4 a brief description of the two-coupling
study of the interface tension with a bootstrap error-analysis
is presented.
Section 5 is devoted to our conclusions.

\section{Simulation}

The lattice action of the SU(2)-Higgs model is given by
\begin{eqnarray}
S[U,\varphi] &=& \beta \sum_{pl}
\left( 1 - \half {\rm Tr\,} U_{pl} \right)\nonumber \\
&&+ \sum_x \left\{ \half{\rm Tr\,}(\varphi_x^+\varphi_x) +
\lambda \left[ \half{\rm Tr\,}(\varphi_x^+\varphi_x) - 1 \right]^2 -
\kappa\sum_{\mu=1}^4
{\rm Tr\,}(\varphi^+_{x+\hat{\mu}}U_{x,\mu}\,\varphi_x)\right\} \: ,
\end{eqnarray}
where $U_{x,\mu}$ denotes the SU(2) gauge link variable and $U_{pl}$
is the path-ordered product of the four $U_{x,\mu}$ around a
plaquette. The symbol $\varphi_x$ stands for the Higgs field, which is
also written as $\varphi_x=\rho_x\cdot\alpha_x$, with $\rho_x \in
{\bf R^+}$ and $\alpha_x \in \mbox{\rm SU(2)}$. Details of the simulation
algorithms can be found in \cite{su2higg_npb,zoltankarl}.

The order parameters under study are $\rho^2$, $\slog$, the density of
$\Slog\equiv S[U,\varphi]-3\sum_x\log(\rho_x)$, and
$L_\varphi$, the density of $\sum_{x\mu} L_{\varphi,x\mu}
\equiv \sum_{x\mu} \half {\rm
Tr\,}(\varphi^+_{x+\hat{\mu}}U_{x,\mu}\,\varphi_x)$. The latter
plays an important r\^ole in our reweighting \cite{ferrenberg}, which
has been used to obtain data in the vicinity of
the simulation point~\cite{su2higg_npb}.

In our finite-temperature simulations we have used elongated lattices:
$V=L_t\times L_{xy}^2\times L_z$, where $L_t=2,\ L_{xy} \ll L_z$.
Keeping the bare parameters $\beta = 8.0$, $\lambda = 3.0\cdot{}10^{-4}$
and $L_t=2$ fixed we tuned $\kappa$ to its critical value.  These
parameters corresponds to a Higgs boson mass of roughly 35 GeV at
$T=0$.

As will be shown in Section 3, the calculation of $\sigma$
by use of the two-coupling method presupposes  a
precise determination of the critical hopping parameter $\kappa_c$
on large lattices. We suggest here a new method, which in
practice gives the same and very precise $\kappa_c$ as the
multicanonical method \cite{neuhaus}. However,
using this method the notoriously large
autocorrelation time of the multicanonical simulation can be
reduced dramatically.

The method is based on the observation that, for sufficiently large
elongated lattices,
a flat regime appears between the peaks in the probability
distribution of an order parameter \cite{su2higg_npb}.
The peaks correspond to pure phase configurations. Between
the peaks the system is dominated by configurations with two interfaces
perpendicular to the long direction $z$, and two bulk phases.
At the transition point the free energy
densities of these bulk
phases are the same, the ratio of the occupied volumes
is arbitrary, thus the distribution is flat. By use of
constrained simulation \cite{bhanot} one can enforce
that an order parameter
of the system stays in a given short interval between
the peaks. Close to the transition point one can use a reweighting
procedure to determine the value of the hopping parameter at which the
distribution is flat.

In practice we have chosen $\slog$ as the order parameter
of the constrained simulation.
Since $\slog$ remains unchanged by the overrelaxation algorithms
\cite{zoltankarl}, only
the heatbath algorithms have to be modified. Every proposal that
leads to expectation values of $\slog$ outside the selected interval,
is rejected. The observed acceptance rate of the
algorithm is larger than 98\%.
The lattice size under study was
$2\times24^2\times256$. In a
series of two short runs without any bounds on $\slog$, the position of
the two peaks, belonging to the pure phases, was
estimated. As a consequence the interval $\slog \in [4.90,4.95]$ has been
chosen to determine $\kappa_c$.
We have performed 18000 measuring sweeps at
$\kappa=0.12865$, and by using
reweighting we have obtained results in the vicinity of this point.
The integrated autocorrelation time turned out to be
$\tau_{\rm int} = 21(2)$ sweeps for $\slog$
and $L_{\varphi,x\mu}$. This value is much smaller than a typical
autocorrelation time for a multicanonical simulation.
The reason is the following.
In the case of a multicanonical simulation the system must completely go from
one phase to the other in order to obtain an independent configuration.
In our
method only the ratio of the bulk phases changes slightly. However, this
is enough to tell whether the free energy densities are the same
or not.

The transition point is defined by the implicit relation for $\kappa_c$
\begin{equation} \label{kapcond}
\langle\slog\rangle_{\kappa_c} \equiv \half \left(\smin+\smax\right)\, ,
\end{equation}
with $\smin$ and $\smax$ denoting the bounds on $\slog$. In our case this
definition is equivalent to the condition that at $\kappa_c$
the distribution is flat in the given interval.

For the determination of the statistical errors on $\langle \slog
\rangle_\kappa$ we  have used the results of a bootstrap procedure with 3000
iterations and 30 independent subsamples (see Section 4).
The final result of this analysis
is $\kappa_c=0.1286565(7)$.
\begin{figure} \begin{center}
\epsfig{file=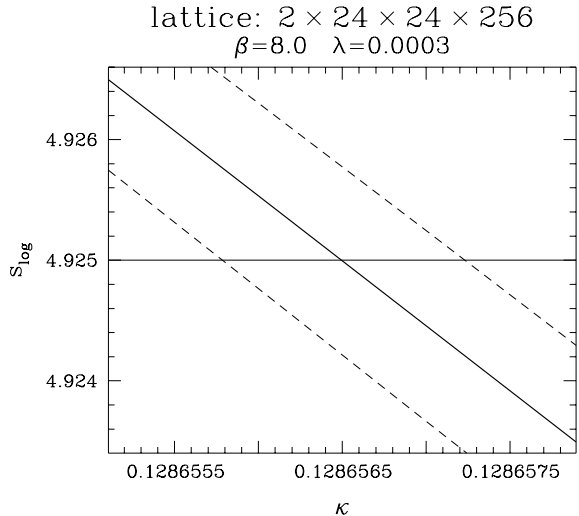,bbllx=40pt,bblly=280pt,bburx=560pt,%
bbury=720pt,width=8.0cm}
\caption{\label{kapcfig} \sl The order parameter $\slog$ as a function
of the hopping parameter.
The dashed lines give the range of the statistical error.
{}From\/ $4.90 < \slog < 4.95$ the result\/ $\kappa_c=0.1286565(7)$ is
obtained. }
\end{center}\end{figure}
Within the error bars,
it agrees with the result from the $2\times 16^2\times256$
lattice (see Section 3).
Thus, the finite-size ef\-fects of $\kappa_c$ beyond
$L_{xy}=24$ are expected to be
small.  The values of $\slog$ as a
function of $\kappa$ are plotted in figure ~\ref{kapcfig}.

The above procedure gives the same $\kappa_c$ for the
upper half and for the lower half of the interval.
This fact can be considered as a check that the system was
in the flat regime.
We frequently tested that the configurations contained
two bulk phases and two interfaces in between.

As a non-trivial cross-check of the method we should mention that,
for $\lambda=0.0001$ ($m_H\approx 18$ GeV)
the results of \cite{su2higg_npb} obtained by the
multicanonical method have been reproduced with this new procedure.

\section{Tunnelling and energy splitting}
The electroweak symmetry is spontaneously broken;
however, this phenomenon does not appear in lattice simulations, since
at finite volume the symmetry is never broken.
In this  case a small energy splitting ($E_0$) appears  between the
ground state and the first excited state, which vanishes
in the infinite volume limit.
The appearance of this $E_0$ at finite volume is due
to the tunnelling effect between the symmetric and the Higgs phase.
In the $z$-direction different domains can be observed, with
domain walls between them. The associated interface energy
determines the tunnelling mass ($E_0$) via
\begin{equation}\label{e0sigma}
E_0 =C \exp(-L_t L_{xy}^2 \sigma).
\end{equation}
This tunnelling effect has been used
to determine, for instance, the interface tension
of the four-dimensional Ising model in the broken phase
\cite{vacuum_1,vacuum_2}, to calculate the
interface tension in three-dimensional systems \cite{munster90},
to determine the interface tension of the
finite-temperature confinement--deconfinement phase transition of the
pure SU(3) gauge theory \cite{tunnel_qcd},
and to determine the interface
tension of the electroweak phase transition \cite{bunk95}.
In general, the prefactor $C$ of eq. (3) can depend on $L_{xy}$; but
in our case the leading contribution is $L_{xy}$-independent.

\begin{table*}[tb] \begin{center}
\begin{tabular}{|c|r|r@{.}l|}
\hline\hline
Lattice&\multicolumn{1}{c|}{sweeps}&\multicolumn{2}{c|}{$\kappa_c$}%
\\ \hline \hline
2\ts4\ts4\ts128& 243200&  0&12882(1)   \\
2\ts8\ts8\ts128& 156000&  0&12869(1)   \\
2\ts12\ts12\ts128& 512000& 0&128663(2)  \\ 
2\ts12\ts12\ts256& 512000& 0&128665(2)  \\ 
2\ts16\ts16\ts128& 448200& 0&128659(1) \\ 
2\ts16\ts16\ts256& 419200& 0&128658(1)  \\ \hline
\end{tabular}
\caption{\label{kappatab} \sl
Critical hopping parameter $\kappa_c$ for various lattice sizes.}
\end{center} \end{table*} %

\begin{figure*}[tb] \begin{center}
\epsfig{file=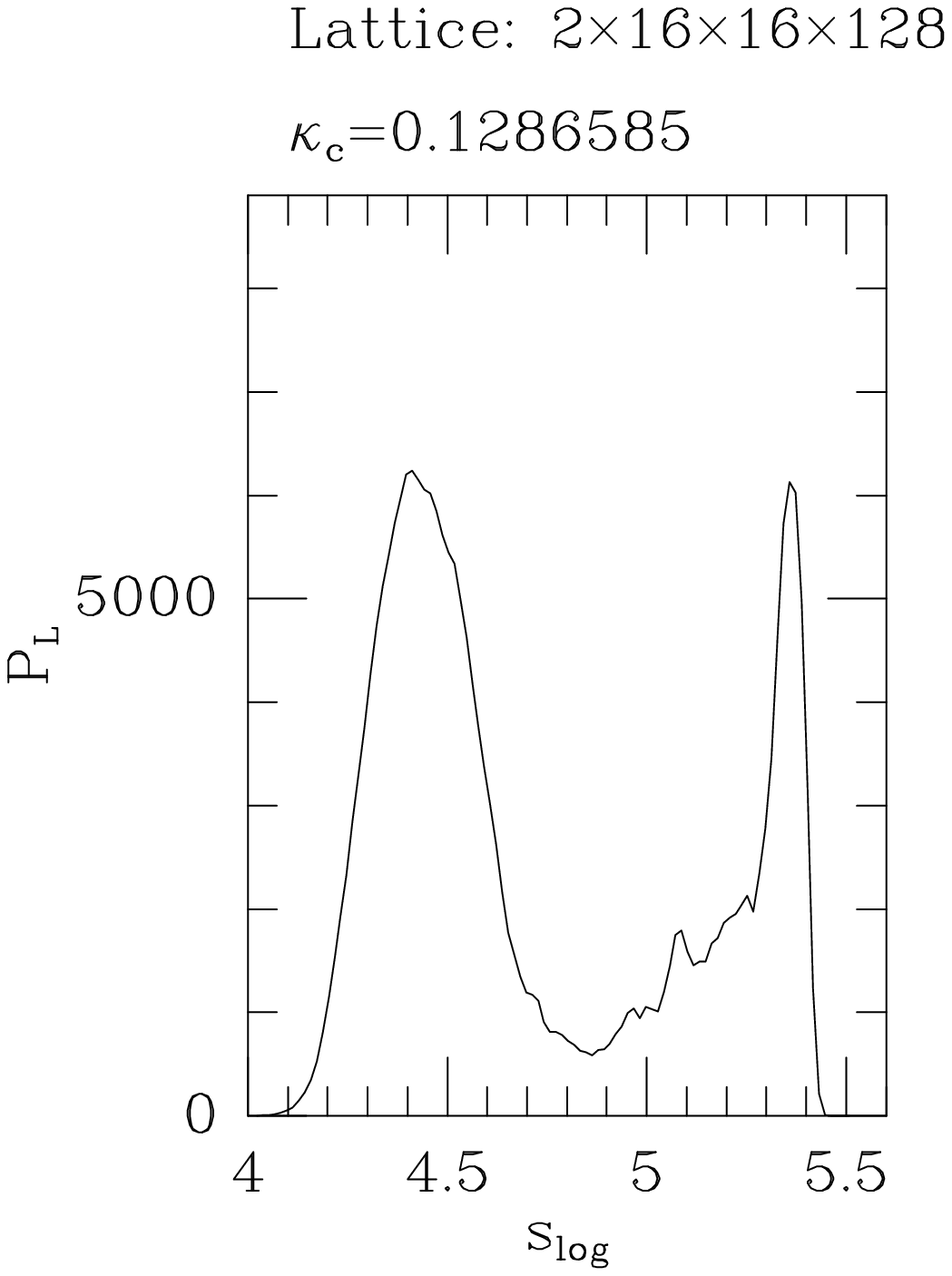,
bbllx=-60pt,bblly=70pt,bburx=470pt,bbury=410pt,width=11cm}
\caption{\label{2peak}\sl Distribution of the order parameter
$\slog$  at the transition point.}
\end{center}
\end{figure*}

According to eq.~(\ref{e0sigma}) we have measured the energy splitting on
dif\-ferent lattice sizes. The number of sweeps and the critical
hopping parameters are listed in table~\ref{kappatab}.
The observed exponential autocorrelation times are  ${\cal
O}(100)$ in the case of smaller volumes ($L_{xy}=4,8$)
and ${\cal O}(1000)$
in the case of larger volumes ($L_{xy} = 12,16$).  For larger volumes, the
clear two-peak structure  with the equal-height criterion for the
$\slog$-distribution has been used to determine the transition point
(cf.~figure ~\ref{2peak}).
In the case of smaller lattices, $\kappa_c$ has been provided by
the peak in the susceptibility.

\begin{figure}\begin{center}
\epsfig{file=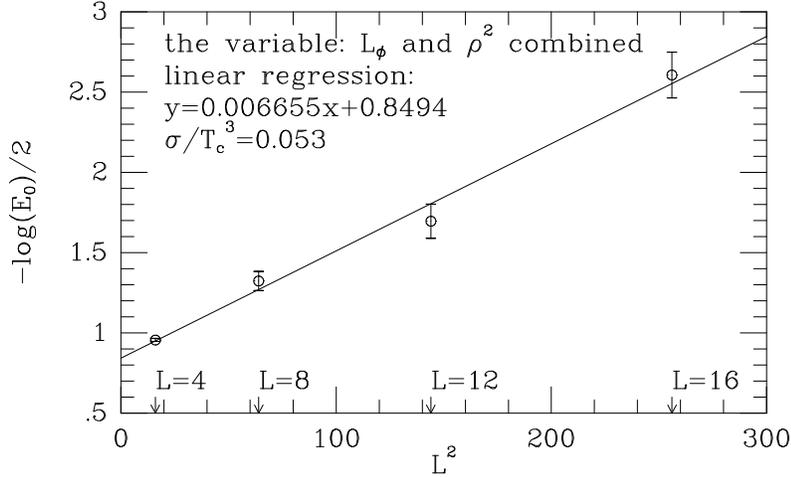,
bbllx=45pt,bblly=400pt,bburx=575pt,bbury=730pt,width=10cm}
\caption{\label{split_sgm}\sl
$-\log(E_0)/L_t$ as a function of $L_{xy}^2$. The slope gives the
interface tension.}
\end{center}\end{figure}

For the determination of the smallest mass in the Higgs channel the
operators $L_{\varphi,x\mu}$ and $\rho_x^2$ have been used at the transition
point.
The correlation functions have been determined in the $z$-direction
with a relative error of ${\cal O}(10^{-2})$.
In the case of smaller lattices ($L_{xy}=4,8$), mass fits
with one exponential can be used, and the determination of the
splitting mass does not cause any problem. However, for $L_{xy}=12,16$
the analysis is more complicated. By use of a one-mass fit we have
determined the masses for dif\-ferent $z$-slice intervals with length of
10 lattice points. Let $m_i$ denote the mass, fitted on the $z$-slice
interval: $i\le z \le i+10$. Since the r\^ole of the larger masses
becomes less and less important
for larger distances one obtains the following condition:
\begin{equation}
{\rm if} \quad i>j\quad \Longrightarrow \quad m_i<m_j.
\end{equation}
For the real data, $m_i$ decreases for a while; then, due to the noise,
it starts to increase.  We have chosen our fit interval in such a way, that
monotony holds. Typical results are 5--25 and 5--55 for
$L_z=128$ and $L_z=256$, respectively. For these intervals we
have performed two-mass fits. Since there are several local minima in the
$\chi ^2$-fit, we take those results, which appear for both
$z$-extensions $L_z=128$ and $L_z=256$.  The statistical errors are
estimated by a jackknife analysis with  256 independent subsamples.
The results are listed in table 2.
The origin of the difference between the $E_0$ values  for $L_{xy}=16$
comes from the following fact.
In the studied region the decay of the $L_\varphi$ correlation
function is dominated by the smallest mass, however, in case
of $\rho^2$ the decay is still dominated by the next excitation
($E_1 \approx 0.1$).
Therefore, it is better to consider $E_0$ from $\rho^2$ as an
upper bound.
Figure ~\ref{split_sgm} contains $-\log(E_0)/L_t$ as a function of
$L_{xy}^2$.  The slope of the fitted line gives the interface tension
\begin{equation}\label{split_res}
\frac{\hat\sigma}{T_c^3}=0.053(5).
\end{equation}
The error contains the statistical error and an estimate
of the systematics. The latter has been obtained from the
difference between the results of the two
dif\-ferent operators $L_{\varphi,x\mu}$ and $\rho_x^2$.
The simulations have been performed on the Quadrics-APE machine of DESY,
and as a consistency check, some of the calculations have been repeated
with 64-bit floating point arithmetics.

The prediction of the perturbative approach \cite{fodor} is
\begin{equation}
{\sigma \over T_c^3}\Bigg|_{perturbative}=0.060(6).\end{equation}
The error displayed here comes from the uncertainties
in the lattice determination of the renormalized masses and coupling
\cite{buchmuller}.

\begin{table*}[tb]\label{masses}
\begin{center}
\begin{tabular}{|c|cccc|}
\hline\hline
$L_{xy}$ & 4 & 8 & 12 & 16 \\
\hline\hline
$E_0$ from $L_{\varphi}$ & 0.147(3) & 0.066(4) & 0.034(1) & 0.0040(6) \\
$E_0$ from $\rho_x^2$   & 0.149(4) & 0.076(4) & 0.033(6) & 0.0069(1) \\ \hline
\end{tabular} \end{center}
\caption{\sl \label{split_tab}%
The measured energy splitting for dif\-ferent lattice extensions.}
\end{table*}

\section{Two-coupling method and bootstrap analysis}

This method is described in ref.~\cite{su2higg_npb}, therefore
we just sketch the basic ideas.

The lattice with $z$-extension $L_z$ much larger than the others, is
divided into two halves \cite{potreb}.
In the part with lower $z$-coordinates we
choose the hopping parameter to be $\kappa_1 < \kappa_c$ and in the
other one to be $\kappa_2 > \kappa_c$. With a suitable initialization
a mixed state with two interfaces is obtained. Let
$L^{(i)}_\varphi(\kappa_1,\kappa_2)$ denote the expectation value of
$L_{\varphi}$ in the part with hopping parameter $\kappa_i$.
One can estimate the interface tension by
\begin{equation} \label{sigmalim}
\sigma \simeq
\frac{1}{2} \,
\lim_{\kappa_2\searrow\kappa_c}\, \lim_{\kappa_1\nearrow\kappa_c}
\left\{\left(\kappa_2-\kappa_1\right)
\lim_{L_z\to\infty}L_z \left[
L^{(1)}_\varphi(\kappa_1,\kappa_2)-L^{(2)}_\varphi(\kappa_1,\kappa_2)
\right]\right\}\, .
\end{equation}
With the ansatz
\begin{equation}\label{c_ansatz}
L^{(i)}_\varphi(\kappa_1,\kappa_2) = - \frac{c_i}{\kappa_i-\kappa_c}
+ b_i  +a_i\cdot (\kappa_i-\kappa_c) + {\cal O}(\kappa_i-\kappa_c)^2
\end{equation}
a finite volume estimator of the interface tension is
obtained from
\begin{equation}\label{sigma_c}
\hat \sigma = L_z\cdot(c_1 + c_2)\, .
\end{equation}
As eq.~(\ref{sigmalim}) illustrates, $\sigma$ depends only on the
dif\-ference $\Delta L_\varphi \equiv [L_\varphi^{(2)}-L_\varphi^{(1)}]$.

In table~\ref{kappapair} the results for $L_\varphi^{(1)}$,
$L_\varphi^{(2)}$ and $\Delta L_\varphi$ on a $2\times24^2\times200$
lattice are listed.
\begin{table}\begin{center}
\begin{tabular}{|ccc|r@{.}lr@{.}lr@{.}l|}
\hline \hline
$\kappa_1$ & $\kappa_2$ & sweeps
&\multicolumn{2}{c}{$L_{\varphi}^{(1)}$}
&\multicolumn{2}{c}{$ L_{\varphi}^{(2)}$}
&\multicolumn{2}{c|}{$\Delta L_{\varphi}$} \\ \hline\hline
0.12845 & 0.12885 & 10000 & 1&0607(30) & 7&9042(91) & 6&8435(83) \\
0.12849 & 0.12881 & 10000 & 1&0614(32) & 7&007(10)  & 5&9461(92) \\
0.12853 & 0.12877 & 10000 & 1&0587(41) & 6&058(12)  & 4&999(10) \\
0.12855 & 0.12875 & 12000 & 1&0684(38) & 5&581(12)  & 4&513(11) \\
0.12857 & 0.12873 & 14000 & 1&0678(42) & 5&035(12)  & 3&967(11) \\
\hline
\end{tabular}
\caption{\label{kappapair} \sl Results for $L_\varphi^{(1)}$,
$L_\varphi^{(2)}$ and $\Delta L_\varphi$ on a $2\times24^2\times200$
lattice.}
\end{center}\end{table}
We estimated the integrated autocorrelation time from the quotient of
true and naive statistical errors. For $L_\varphi^{(1)}$ and
$L_\varphi^{(2)}$ the autocorrelation time
is always found to be smaller than
50~sweeps. The errors, quoted in the table, are obtained by binning.

The statistical errors of $\Delta L_\varphi$ are always smaller than
those of $L_\varphi^{(2)}$. This is due to a kind of shift
of the interfaces during the simulation; for instance, in  the case of an
interface moving from the region with $\kappa_2$ into the one with
$\kappa_1$, both $L_\varphi^{(i)}$ will increase, but the dif\-ference
between them is not that much af\-fected.
Thus, there is a strong correlation between the two
$L_\varphi^{(i)}$-values obtained on the same configuration. This has
to be taken into account when estimating statistical errors for
the interface tension, and is achieved by a bootstrap analysis.

The bootstrap procedure has been discussed in \cite{efron} and used
e.g.~in \cite{gupta}. We shortly discuss our analysis.  Consider a set of
$N$ independent measurements for some primary quantities from a single
Monte Carlo run. To form a bootstrap sample from the original $N$
measurements we take $N$ objects of them with repetition, thus some
may be chosen a couple of times. The number of bootstrap samples
is huge, $N^N$, therefore one randomly generates representatives.
Since correlations between the different primary quantities should
be considered, these quantities have to
emerge from the same Monte Carlo configurations for a given bootstrap sample.
One computes in each
sample a given secondary quantity $a$ from the sample averages of the primary
quantities (e.g.~mass fits from correlation functions).
An estimate of
the statistical error is given by the dispersion of the $a$-values,
thus by the two values left and right to the median containing 68.3\%
(34.15\% in both directions) of the distribution.

In order to eliminate the influence of autocorrelation we have
combined 200 measurements to bins for each $\kappa$-pair. (E.g.~for
the first row of table~\ref{kappapair} the bin averages provide $N=50$
independent values for the quantities $L_\varphi^{(1)}$ and
$L_\varphi^{(2)}$.)  Our data are not obtained in a single Monte Carlo
run but in 5 different runs for different $\kappa$-pairs.  For each
$\kappa$-pair we generated a bootstrap sample and for the obtained
samples we determined $L_\varphi^{(1)}$ and $L_\varphi^{(2)}$ as a
function of $\kappa$. Using the form of eq.~(\ref{c_ansatz}) we
performed for these functions two $\chi^2$-fits. This procedure has
been repeated 10000 times. The statistical error has been determined
from the distribution of the $\hat\sigma$.

%

\begin{figure}\begin{center}
\epsfig{file=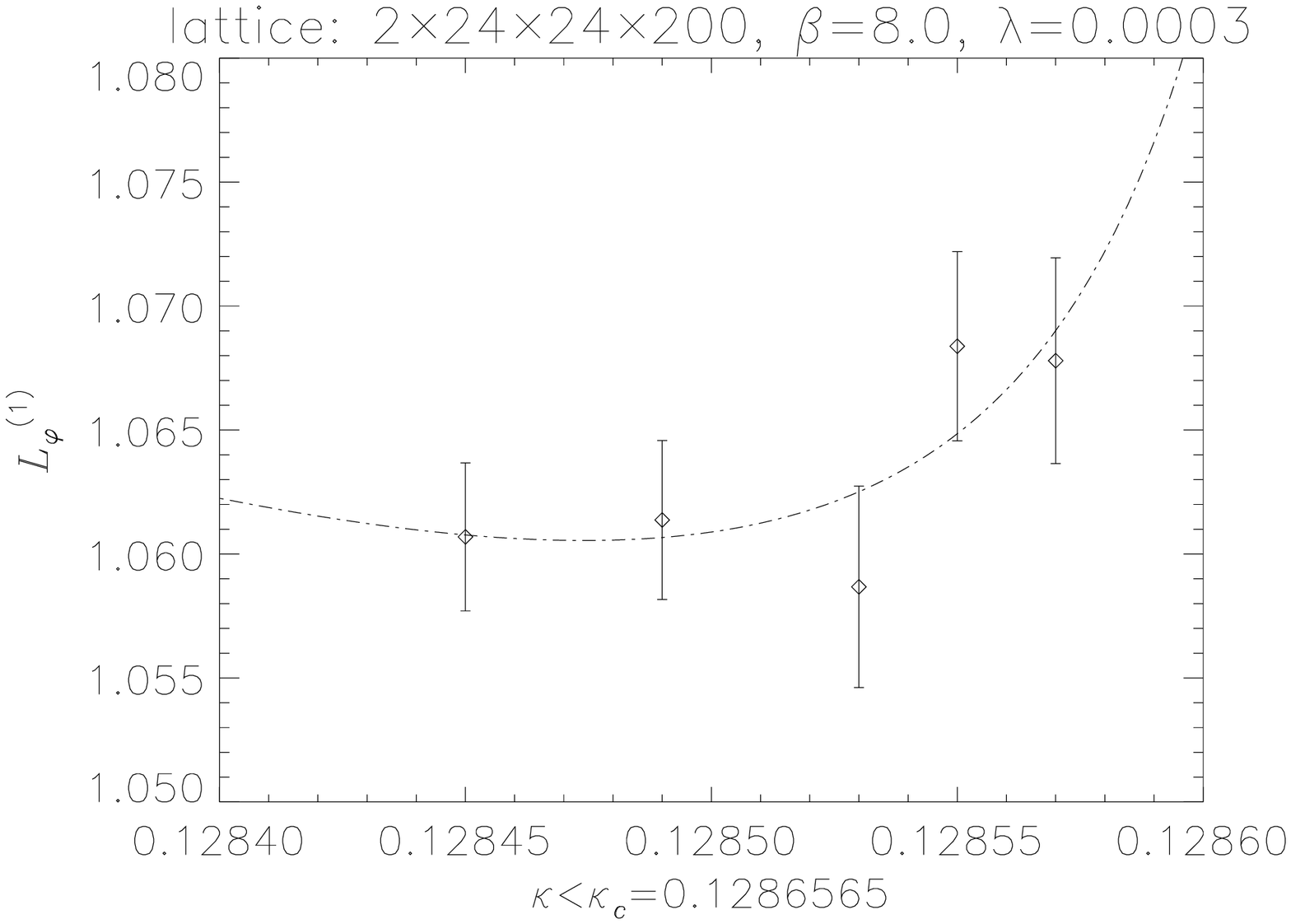,width=8cm}
\epsfig{file=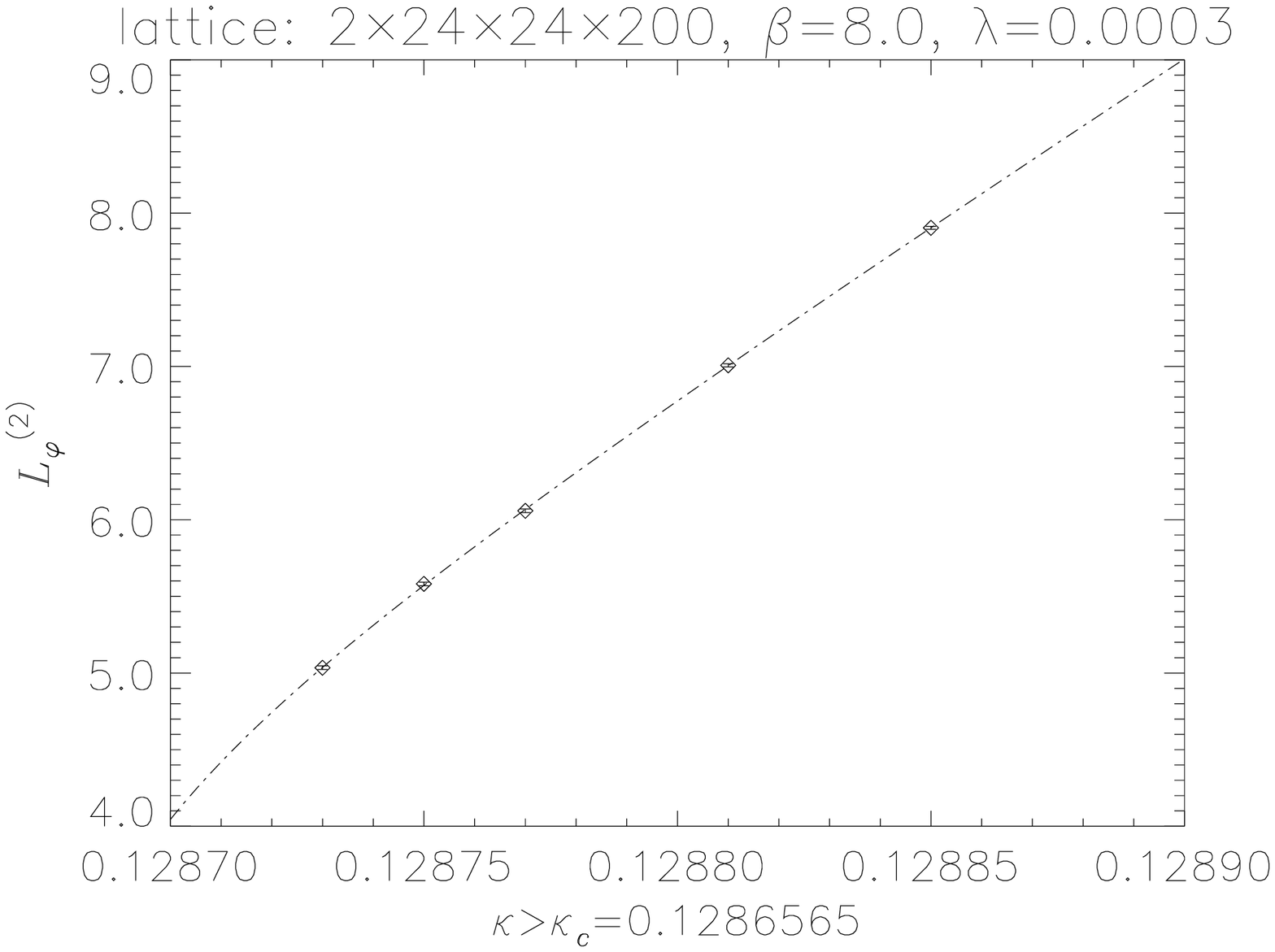,width=8cm}
\caption{\label{lphiplot}\sl Least $\chi^2$-fits for the determination
of $\hat\sigma$ from the two-coupling method. The $\chi^2$-s are
$\chi_1^2=1.88$ and $\chi_2^2=1.50$, respectively.}
\end{center}\end{figure}

Figure ~\ref{lphiplot} contains the combined data points with the
corresponding fits. This yields
\begin{equation}\label{twok_res}
\frac{\hat\sigma}{T_c^3} = 0.065(9+1)\, .
\end{equation}
The first number in the parenthesis is the statistical error, the
second one stands for the uncertainty in the critical hopping parameter. For
the individual fits we have $c_1=0.00000277(290+5)$ and
$c_2=0.0000378(61+7)$
with $\chi_1^2=1.88$ and $\chi_2^2=1.50$, respectively.
The error on $(L_z T_c^{-3}\cdot c_2)$ is even larger than that of
$\hat\sigma/T_c^3$. This is an indication that the correlations have been
taken into account. In order to check our error estimate based on the
bootstrap procedure, we have also performed a jackknife error analysis with
80 independent jackknife samples. The results of the two different methods
are in complete agreement.

In order to check whether the measured curvature of
$L^{(i)}_\varphi(\kappa_1,\kappa_2)$ as a function of $\kappa_i$ is
really described by  eq.~(\ref{c_ansatz}),
we have also tried a 3-parameter quadratic ansatz. The results are
$\chi^2_1=1.85$ and $\chi^2_2=4.15$. The value of $\chi^2_2$ shows
that the data cannot be well described by a
quadratic fit.
Due to the large error bars in the symmetric phase
(cf.~figure ~\ref{lphiplot}) $\chi^2_1$ is not very sensitive to the proposed
fit-function.
A 4-parameter fit with a quadratic term in addition to eq.~(\ref{c_ansatz})
results in $\chi_1^2=1.85$ and $\chi_2^2=1.49$.
Since the $\chi^2$-s are nearly unchanged, the fourth parameter
does not seem to represent a new degree of freedom and the value
for $\hat\sigma /T_c^{3}=0.063$ does not deviate significantly from
that of eq. (\ref{twok_res}).

It is also possible to determine $\hat\sigma$ by using the
$\Delta L_\varphi$-values of table 3 together with the
rough estimate $\kappa_c=0.12865$ for the transition point,
which has been obtained by the two-coupling method.
One gets $\hat\sigma/T_c^3=0.075(11)$
with $\chi^2 = 0.77$. In this case the error comes from 1000 normally
distributed random data and the uncertainty in $\kappa_c$ has not been
considered.
With individual fits for both $L^{(i)}_\varphi$
as previously described, the result is $\hat\sigma/T_c^3=0.075(11)$ for
$\kappa_c=0.12865$ with $\chi^2_1=1.89$ and $\chi^2_2=1.54$.
{}From this agreement we conclude that the correlations
between the $L_\varphi^{(i)}$-values are treated similarly in both methods.
However, the difference between this result and that of eq. (\ref{twok_res})
emphasizes the importance of a precise $\kappa_c$-determination.

\section{Conclusions}

We have determined the interface tension of the SU(2)-Higgs model with two
dif\-ferent methods. Within the error bars
the results of eqns.~(\ref{split_res}) and
(\ref{twok_res}) agree with each other and with the prediction of the
perturbation theory. When comparing the errors, it
should be kept in mind that the product of the lattice volume and the number of
sweeps (thus, CPU-time) is an order of magnitude smaller in the case of the
two-coupling method.

Together with the results of ref.~\cite{su2higg_npb},
in which Binder's method \cite{binder} has been used with multicanonical
simulation, in total three
dif\/ferent approaches for the measurement of the interface tension were
studied. No significant difference between the results of
these methods has been
observed. The two-coupling method
turned out to be the most economic way to estimate the interface
tension.

\subsection*{Acknowledgements}
Helpful discussions with  K. Jansen, A. Jaster, I.
Montvay, G. M\"unster and R. Sommer are acknowledged.

Most of the energy splitting simulations
have been carried out on the Quadrics-APE Computers of DESY-IfH in
Zeuthen, Germany. The rest was done on the CRAY
Y-MP8/864 of HLRZ in J\"ulich, Germany.

Two of us (F. Cs. and Z. F.) have been partially supported by Hungarian
Science
Grant under Contract No. F1041/3-2190.

\vfill\eject


\begin{thebibliography}{99}
\bibitem{kuzmin}
V. A. Kuzmin, V. A. Rubakov and M. E. Shaposhnikov, Phys. Lett. B155
(1985) 36.
\bibitem{arnold}
P. Arnold and O. Espinosa, Phys. Rev. D47 (1993) 3546;\\
W. Buchm\"uller, Z. Fodor, T. Helbig and D. Walliser,
Ann. Phys. 234 (1994) 260.
\bibitem{fodor}
Z. Fodor and A. Hebecker, Nucl. Phys. B432 (1994) 127.
\bibitem{buchmuller}
W. Buchm\"uller, Z. Fodor, A. Hebecker, DESY-95-028, hep-ph/9502321,
to app. in Nucl. Phys. B.
\bibitem{bunk}
B. Bunk, E.-M. Ilgenfritz, J. Kripfganz and A. Schiller,
Nucl. Phys. B403 (1993) 453.
\bibitem{kajantie}
K. Kajantie, K. Rummukainen, M. Shaposhnikov, Nucl. Phys. B407 (1993) 356;
K. Farakos, K. Kajantie, K. Rummukainen, M. Shaposhnikov,
Phys. Lett. B 336 (1994) 494.
\bibitem{su2higg_phylett}
F. Csikor, Z. Fodor, J. Hein, K. Jansen, A. Jaster, I. Montvay,
Phys. Lett. B334 (1994) 405.
\bibitem{su2higg_npb}
Z. Fodor, J. Hein, K. Jansen, A. Jaster and I. Montvay,
Nucl. Phys. B439 (1995) 147.
\bibitem{karsch}
F. Karsch, T. Neuhaus and A. Patk\'os, Nucl. Phys. B441 (1995) 629.
\bibitem{zoltankarl}
Z.~Fodor and K.~Jansen, Phys.~Lett. B331 (1994) 119.
\bibitem{ferrenberg}
A.M.~Ferrenberg, R.H.~Swendsen, Phys.~Rev.~Lett. {61} (1988) 2635;
Phys.~Rev.~Lett. {63} (1989) 1195.
\bibitem{bhanot}
G. Bhanot, S. Black, P. Carter and R. Salvador,
Phys. Lett. B183 (1987) 331.
\bibitem{neuhaus}
B. A. Berg and T. Neuhaus, Phys. Rev. Lett. 68 (1992) 9.
\bibitem{vacuum_1} K.~Jansen, J. Jersak, I. Montvay,
G. M\"unster, T. Trappenberg, U. Wolff,
Phys.~Lett. B213 (1988) 203.
\bibitem{vacuum_2} K.~Jansen, I. Montvay, G. M\"unster, T. Trappenberg,
U. Wolff,
Nucl.~Phys. B322 (1989) 698.
\bibitem{munster90}
G. M\"unster, Nucl. Phys. B340 (1990) 559; S. Klessinger, G. M\"unster,
Nucl. Phys. B386 (1992) 701.
\bibitem{tunnel_qcd} B.~Grossmann et al.,
Nucl.~Phys. B396 (1993) 584.
\bibitem{bunk95}
B. Bunk, Nucl. Phys. B (Proc. Suppl.) 42 (1995) 566.
\bibitem{potreb}
J. Potvin and C. Rebbi, Phys. Rev. Lett. 62 (1989) 3062;
S. Huang, J. Potvin, C. Rebbi and S. Sanielevici,
Phys. Rev. D42 (1990) 2864,
errata: D43 (1991) 2056.
\bibitem{efron}
B.\ Efron, SIAM\ Review\ 21 (1979) 460.
\bibitem{gupta}
R. Gupta et al. Phys. Rev. D36 (1987) 2813.
\bibitem{binder}
K.~Binder, Z.~Phys. B43 (1981) 119;
 Phys.~Rev. A25 (1982) 1699.
\end{thebibliography}
\end{document}